\documentclass[review,usenatbib]{elsarticle}

\usepackage{lineno,hyperref}
\usepackage[american]{babel}
\usepackage[toc,page]{appendix}
\usepackage{amsmath}
\usepackage{amssymb}
\usepackage{graphicx}
\usepackage{subcaption}
\usepackage{listings}
\usepackage{float}
\usepackage{aas_macros}
\usepackage[left=1cm, right=1cm, top=2cm, bottom=1.5cm]{geometry}
\usepackage{spverbatim}
\usepackage{color, soul}

\newcommand{\StePS}{\texttt{StePS} }

\journal{Astronomy and Computing}

\begin{document}
\twocolumn[{%
\begin{frontmatter}
\title{StePS: A Multi-GPU Cosmological N-body Code for Compactified Simulations}

\author[elteaddress]{G\'abor R\'acz}
\ead{ragraat@caesar.elte.hu}

\author[hawaiiaddress]{Istv\'an Szapudi}

\author[elteaddress]{L\'aszl\'o Dobos}

\author[elteaddress]{Istv\'an Csabai}

\author[baltimoreaddress]{Alexander S. Szalay}

\address[elteaddress]{Department of Physics of Complex Systems, E\"{o}tv\"{o}s Lor\'and University, Pf. 32, H-1518 Budapest, Hungary}
\address[hawaiiaddress]{Institute for Astronomy, University of Hawaii, 2680 Woodlawn Drive, Honolulu, HI, 96822}
\address[baltimoreaddress]{Department of Physics and Astronomy and Department of Computer Science, Johns Hopkins University, 3400 N. Charles Street, Baltimore, MD 21218}

\begin{abstract}
	We present the multi-GPU realization of the \StePS (Stereographically Projected Cosmological Simulations) algorithm with MPI-OpenMP-CUDA hybrid parallelization and nearly ideal scale-out to multiple compute nodes. Our new zoom-in cosmological direct N-body simulation method simulates the infinite universe with unprecedented dynamic range for a given amount of memory and, in contrast to traditional periodic simulations, its fundamental geometry and topology match observations. By using a spherical geometry instead of periodic boundary conditions, and gradually decreasing the mass resolution with radius, our code is capable of running simulations with a few gigaparsecs in diameter and with a mass resolution of $\sim 10^{9}M_{\odot}$ in the center in four days on three compute nodes with four GTX 1080Ti GPUs in each. The code can also be used to run extremely fast simulations with reasonable resolution for fitting cosmological parameters. These simulations are useful for prediction needs of large surveys. The \StePS code is publicly available for the research community.
\end{abstract}


\end{frontmatter}
}]

\section{Introduction}

The evolution of the dark matter structures in an expanding universe is usually solved by the N-body method\cite{2005MNRAS.364.1105S, 1996ApJ...462..563N}. In this approximation the density of the ideal dark matter fluid is sampled by a finite number of smoothed tracer particles and the only interaction between dark-matter particles is Newtonian gravity. Cosmological N-body simulations play an important role in understanding the structure formation of dark matter in the non-linear regime. N-body simulations allow for testing cosmological models and fitting cosmological parameters by following the evolution of the power spectrum $P(k)$, the angular power spectrum $C_l(r)$ and the halo mass function. In a last 44 years, these simulations have gone trough great improvements: from the first simulations that had only $10^3$ bodies\cite{1974ApJ...187..425P}, nowadays it is possible to run simulations with 8 trillion dark matter particles\cite{2017ComAC...4....2P}. This speed up is achieved by faster computers and by algorithmic improvements such as the use of tree-algorithms\cite{1985SJSSC...6...85A, 1986Natur.324..446B} and particle mesh methods\cite{1983MNRAS.204..891K, 1996ApJ...462..563N}.

Most of these simulations are being run in a finite cubic volume with periodic boundary condition which is not supported by observations and causes distortions in the gravitational force field. Our \StePS algorithm eliminates the need for these artificial boundaries, and can simulate an infinite Universe with a topology that matches the observations\cite{2018MNRAS.477.1949R}. The \StePS approach can achieve unprecedented dynamic range by using a small number of particles and a unique isotropic zoom-in method involving the compactification of the spatial extent of the Universe. The relatively small number of particles makes the use of direct force summation possible with low memory needs. The approach is ideal for a relatively simple and very effective GPU parallelization. We demonstrated the effectiveness of the \StePS method in our previous paper\cite{2018MNRAS.477.1949R}.

Modern cosmological simulations rarely use direct force calculation due to its high computation needs. On the other hand, this approach is prevalent where the three-body interactions are significant, such as the globular cluster simulations. Since we focus on cosmological large structure, we only mention here the direct N-body GPU codes NBODY6-GPU \cite{2012MNRAS.424..545N}, NBODY6++GPU \cite{10.1093/mnras/stv817} and $\varphi$-GPU \cite{2011hpc..conf....8B}, interested readers may find further references therein.

The structure of this paper is the following. We present the detailed simulation algorithm with a new multi-GPU parallelization in Section~2. In Section~3 we show how one can generate initial conditions for the \StePS simulation code. Section~4 describes the example simulations that we run to demonstrate the effectiveness of our code by repeating a compactified version of the original Millennium Simulation\cite{2005Natur.435..629S}.

\section{Algorithm}

\subsection{Compactified cosmological simulations}

Computers with finite memory and processing power make it impossible to simulate the infinite universe with constant resolution. Traditional cosmological simulations solve this problem by using periodic boundary conditions which essentially means that the infinite universe is tiled by exactly identical cubic volumes that are repeated in an simple, infinite cubic grid. While these simulations have translational symmetry, they lack rotational symmetry due to the characteristic directions of the grid.

Another way of running a cosmological simulation is to abandon translational symmetry in favor of rotational symmetry. The \StePS algorithm is built on the idea that an infinite universe can be represented in a finite volume using space compactification. In  \cite{2018MNRAS.477.1949R}, we published the details of this algorithm. We repeat the principal ideas and equations in this and the following subsection for convenience only. The original \StePS algorithm uses the inverse 3 dimensional stereographic projection to compactify the infinite space onto the surface of a hypersphere. The stereographic projection can be substituted with any compactification method that conserves rotational symmetry. Compactification is essentially equivalent to re-scaling space in the radial direction around an arbitrarily chosen point while gradually decreasing the mass resolution with distance from the center. This is very similar to zoom-in simulations \cite{1994MNRAS.267..401N, 2003MNRAS.338...14P} except that the resolution changes continuously and smoothly. Computation of the force acting between particles is more complicated in the compact space than in decompactified Cartesian coordinates, therefore, in order to make simulations significantly faster, we transform the constant resolution compact space back into real space. The surroundings of the singularity of the spherical projection at the pole is mapped into an infinite region in real space which is taken into account in the form of an effective radial force pointing outwards that depends on the distance from the center and average density, c.f. Eq.~11 of \cite{2018MNRAS.477.1949R}.

\subsection{Basic Equations}

The expansion of the Universe is described by the Friedmann equations. The first equation can be written as
\begin{equation}
	\left(\frac{\dot{a}}{a}\right)^2 = H_0^2 \cdot \left( \Omega_m \cdot a^{-3} + \Omega_r \cdot a^{-4} + \Omega_\Lambda + \Omega_k\cdot a^{-2} \right),
\label{eq:friedmann}
\end{equation}
where $a(t)$ is the scale factor, $H_0$ is the Hubble constant, and $H=\dot{a}/a$ is the Hubble-parameter. The $\Omega$ density parameters are defined by the ratio of the component energy-density to the critical density. Here we use the present day values. The dimensionless density parameters are the following: $\Omega_m$ is the non-relativistic matter density, $\Omega_\Lambda$ is the dark energy density, $\Omega_r$ is the radiation energy-density and $\Omega_k$ is the spatial curvature density. 

The \StePS code implements N-body cosmological simulations in three different settings. Spherically compactified simulations in comoving or proper coordinates and periodic simulations in comoving coordinates. Below, we derive the basic equations for both periodic and spherical simulation methods.

\subsubsection{Traditional Periodic Simulations}

The equations of motion in the Newtonian approximation, in comoving coordinates are
\begin{equation}
m_i\ddot{\mathbf{x}}_i = \sum\limits_{j=1; j \neq i}^{N} \frac{m_im_j\mathbf{F}(\mathbf{x}_i-\mathbf{x}_j, h_i+h_j)}{a(t)^{3}} - 2 \cdot m_i \cdot \frac{\dot{a}(t)}{a(t)} \cdot \dot{\mathbf{x}}_i,
\label{eq:Comoving_periodic_newtonian}
\end{equation}
where $\mathbf{x}_i$ and $m_i$ are the comoving coordinates and the masses of the particles, whereas $h_i$ and $h_j$ are the softening lengths associated with the particles. The function $m_im_j\mathbf{F}(\mathbf{x}_i-\mathbf{x}_j, h_i+h_j)$ is the magnitude of the force between particles $i$ and $j$, and it depends on the softening kernel and the boundary conditions arising from the periodicity of the simulation box. The \StePS code uses the spline kernel\cite{2005MNRAS.364.1105S,1985A&A...149..135M} for gravitational softening. For zero boundary conditions, $\mathbf{F}(\mathbf{x}, h)$ is given by

\begin{equation}
        \mathbf{F}(\mathbf{x}, h) = -G\mathcal{F}(|\mathbf{x}|, h)\frac{\mathbf{x}}{|\mathbf{x}|},
\label{eq:Force}
\end{equation}
where $G$ is the gravitational constant, and $\mathcal{F}(r, h)$ is
\begin{equation}
        \mathcal{F}(r, h) = \left\{
                \begin{array}{l l}
			\frac{32{r}^{4}}{{h}^{6}}-\frac{38.4{r}^{3}}{{h}^{5}} +\frac{32r}{3{h}^{3}} & \text{\small{if $r < \frac{h}{2}$}}\\
                 \ &\ \\
			-\frac{32{r}^{4}}{3{h}^{6}}+\frac{38.4{r}^{3}}{{h}^{5}}-\\-\frac{48{r}^{2}}{{h}^{4}}+\frac{64r}{3\,{h}^{3}}-\frac{1}{15{r}^{2}} & \text{\small{if  $\frac{h}{2}<r<h$}}\\
                \ &\ \\
                \frac{1}{{r}^{2}} & \text{\small{if $h<r$}.}
                \end{array} \right.
\label{eq:Force_spline_kernel}
\end{equation}
The softening length is set at the beginning of the simulation and, for a constant spatial resolution case, it is the same for every particle.

In this periodic case, multiple images of the particles are taken into account Ewald summation \cite{1921AnP...369..253E, 1991ApJS...75..231H} with the formula
\begin{equation}
        \mathbf{F}(\mathbf{x}, h) = \sum\limits_{\mathbf{n}}-G\mathcal{F}(|\mathbf{x} - \mathbf{n}L|, h)\frac{\mathbf{x}-\mathbf{n}L}{|\mathbf{x}-\mathbf{n}L|},
\label{eq:ForcePeriodic}
\end{equation}
where $L$ is the linear size of the periodic box, and $\mathbf{n}=(n_1,n_2,n_3)$ extends over all integer triplets, in theory up to infinity. A numerical code cannot sum for all integer triplets, so a cut in $\mathbf{n}$ is required. Our code uses the following cut in $\mathbf{n}$: the only valid triplets are where $|\mathbf{x}-\mathbf{n}L| < 2.6L$ is fulfilled\cite{1991ApJS...75..231H}. It is also possible to use quasi-periodic boundary conditions. In this case, only the leading term of the sum in Eq.~\ref{eq:ForcePeriodic} is kept for each pair of particles. 

If the simulation has constant mass resolution everywhere in the periodic box, then the $m_i$ particle masses are calculated directly from the cosmological parameters as
\begin{equation}
	m_i = \frac{\rho_{crit}\cdot\Omega_m\cdot V_{sim}}{N} = \frac{3\cdot H_0^2\cdot\Omega_m}{8\pi G}\cdot \frac{V_{sim}}{N},
\label{eq:Const_part_masses}
\end{equation}
where $V_{sim}$ is the simulation volume, and $N$ is the number of the particles.

\subsubsection{Spherical Simulations}

In the case of spherical zoom-in simulations, the average particle separation and the masses of the particles increase outwards, hence the outer particles represent larger volumes with lower spatial resolution. Eq.~\ref{eq:Const_part_masses} can no longer be used to calculate the masses but the assumption that, at the largest scales, the universe is homogeneous must be kept, and the total mass of the particles must be consistent with the cosmological parameters. The details of particle mass calculation and initial condition generation will be described in Section~\ref{sec:ICs} below. For the spherical geometry, we only set the softening length for smallest-mass particle, and for the rest of the particles the code calculates 
\begin{equation}
        h_i = \sqrt[3]{\frac{m_i}{m_{min}}} \cdot h_{\text{min}},
        \label{eq:softeninglength}
\end{equation}
where $h_{min}$ is the softening length and $m_{min}$ is the mass of the smallest-mass particle. According to this formula, the average density inside a radius of $h_i$ around every particle will be the same. 

As it was shown \cite{2018MNRAS.477.1949R}, the equations of motion in comoving coordinates with non-periodic and isotopic boundary conditions can be derived from Newton's shell theorem. The result is
\begin{equation}
\ddot{\mathbf{x}}_i = \sum\limits_{j=1; j \neq i}^{N} \frac{m_j\mathbf{F}(\mathbf{x}_i-\mathbf{x}_j, h_i+h_j)}{a(t)^{3}} - 2 \cdot  \frac{\dot{a}(t)}{a(t)} \cdot \dot{\mathbf{x}}_i + \frac{4 \pi G }{3}\overline{\rho}\mathbf{x}_i,
\label{eq:Comoving_Spherical_newtonian}
\end{equation}
where $\overline{\rho} = \rho_{crit}\cdot\Omega_m$ is the average matter density and $\mathbf{F}(\mathbf{x}, h)$ is calculated with the spline kernel by using Eq.~\ref{eq:Force}. The last term of the right hand side of the equation is the effective radial force coming from the homogeneous boundary condition.

The \StePS code also can run cosmological simulations with static, non-comoving coordinates in a fully Newtonian way. For more detailed discussion, see \cite{2018MNRAS.477.1949R}.

\subsection{Time integration}

The most time-consuming part of integrating an N-body system is the calculation of the forces, especially if the forces are calculated pairwise. It is vital to minimize the number of force calculations per time step if $N$ is large. For integrating the equations of motion, we used the kick-drift-kick (KDK) leapfrog integrator\cite{1997astro.ph.10043Q}. This is a second order method, yet it needs only one gravitational force evaluation per time step. The integrator uses two different operators, the
\begin{equation}
	K_i(\Delta t): \left\{
                \begin{array}{l l}
			\mathbf{x}_i \longmapsto \mathbf{x}_i\\
                \ &\ \\
			\mathbf{v}_i \longmapsto \mathbf{v}_i + \mathbf{A}_i \cdot \Delta t
                \end{array} \right.
	\label{eq:kick}
\end{equation}
'kick' and the
\begin{equation}
        D_i(\Delta t): \left\{
                \begin{array}{l l}
			\mathbf{x}_i \longmapsto \mathbf{x}_i + \mathbf{v}_i \cdot \Delta t\\
                \ &\ \\
                        \mathbf{v}_i \longmapsto \mathbf{v}_i
                \end{array} \right.
        \label{eq:drift}
\end{equation}
'drift' operator, where $\mathbf{A}_i$ is the acceleration of the particle. The KDK integrator uses two 'kick' and one 'drift' operation per time step, so the time evolution operator is
\begin{equation}
	\tilde{U}(\Delta t) = K\left(\frac{\Delta t}{2}\right) D(\Delta t) K\left(\frac{\Delta t}{2}\right).
\label{eq:KDK}
\end{equation}

Adaptive time steps of the KDK integrator are determined by the time step criterion
\begin{equation}
	\Delta t = \min\left[\Delta t_{\text{max}}, \sqrt{\frac{2\eta_i \epsilon}{|\mathbf{A}_i|}}\right],
	\label{eq:timestepcriterion}
\end{equation}
where $\eta_i=2.8\cdot h_i$ is the Plummer equivalent softening length, $\epsilon$ is the accuracy parameter, and $|\mathbf{A}_i|$ is the acceleration of the particle. $\Delta t_{\text{max}}$ is the maximal allowed length for a time step. The same formula is used in the cosmological code GADGET-2\cite{2005MNRAS.364.1105S}.

Most cosmological N-body codes use the scale factor $a(t)$ instead of cosmic time $t$ as the time variable and apply the formula
\begin{equation}
	\Delta t = \left(a \cdot H_0 \cdot \sqrt{\left( \Omega_m \cdot a^{-3} + \Omega_r \cdot a^{-4} + \Omega_\Lambda + \Omega_k\cdot a^{-2} \right)} \right) ^{-1} \cdot\Delta a
	\label{eq:scalefatortime}
\end{equation}
to calculate the physical time when particle positions are updated. This can be seen as a first-order Euler integration of Eq.~\ref{eq:friedmann}. When integrating the equations of motion in proper coordinates, however, the scale factor does not appear and one has to use $t$ as the time parameter. The \StePS code always uses $t$ as the time parameter for integration, hence, to achieve higher precision when integrating in comoving coordinates, we compute $a(t)$ with the fourth-order Runge--Kutta method with the same time step length that the N-body integrator uses.

\subsection{Force calculation and parallelization methods}

The most time-consuming part of a direct N-body code is the gravitational force calculation, since the execution time scales as $N(N-1)/2$. Every other part of our code scales with $N$ or better, so it is enough to parallelize the force calculation part of the program. The \StePS approach makes a massively parallel implementation possible which can scale out to a large CPU or GPU cluster.

The use of direct force calculation is feasible because, compared to the traditional approach, simulations with radially decreasing resolution have a relatively small number of particles, even when simulating an extremely large volume with high resolution at the center. The small number of the particles carries another advantage: only a few hundred MBs of memory needed to store all the particle data. GPUs are ideal candidates for this type of calculation: they are $\sim 100$ times faster than similarly priced CPUs, and have enough memory to store all particle coordinates and masses.

\begin{figure}
    \centering
        \includegraphics[width=0.48\textwidth]{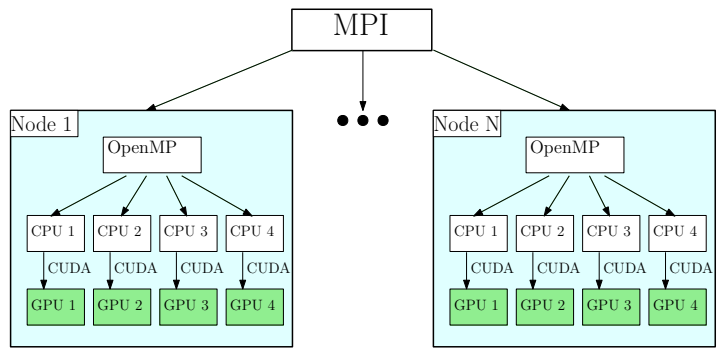}
    \caption{Force calculation in a GPU cluser with MPI-OpenMP-CUDA hybrid parallelization.}
\label{fig:MPI_OMP_CUDA}
\end{figure}

For force calculation, the code allows for two or three levels of parallelization. The first level is the communication between the nodes in the computing cluster. For this, we used the Open Message Passing Interface\footnote{\url{https://www.open-mpi.org/}} (OpenMPI) library. At the first level of the parallelization, the ``main'' node broadcasts particle coordinates and masses to every other node for force calculation. At the second level, the Open Multi-Processing\footnote{\url{https://www.openmp.org/}} (OpenMP) library is used to start as many processing threads as many GPU devices or available or, if no GPUs are used, as many processing cores are available on the node. If only CPUs are used for the force calculation, this is the last level of parallelization. When available, the third level of parallelization is done on the GPUs, implemented with CUDA\footnote{\url{https://developer.nvidia.com/cuda-zone}}. Every node is responsible for calculating $floor(N/N_{node})$ force vectors, except the ``main'' node, which calculates $\textrm{floor}(N/N_{\textrm{node}}) + \textrm{mod}(N,N_{\textrm{node}})$. At the third level of parallelization, every second-level thread is used to manage the corresponding GPU of the node: they copy the particle data to the GPU, wait for the end of the force calculation and copy the calculated forces into the main memory of the node. After force calculation, the ``main'' node collects the calculated force vectors from every node, and does all other calculations, such as integration of the equations of motion and the Friedmann equation. Snapshoting and redshift cone calculation also happens at the ``main'' node.

\subsection{Performance analysis}

\begin{figure}[h!]
    \centering
         \includegraphics[width=0.45\textwidth]{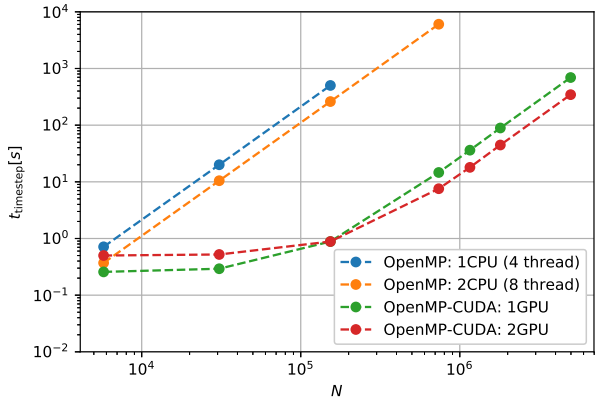}
        \caption{Wall-clock times for one simulation timestep of the \StePS code with OpenMP (CPU-only), and with OpenMP-CUDA parallelization. This benchmark was run on Intel Xeon E5520 CPUs and Nvidia GeForce GTX 1080 Ti GPUs.}\label{fig:CPU_vs_GPU}
\end{figure}

\begin{figure}[h!]
    \centering
         \includegraphics[width=0.45\textwidth]{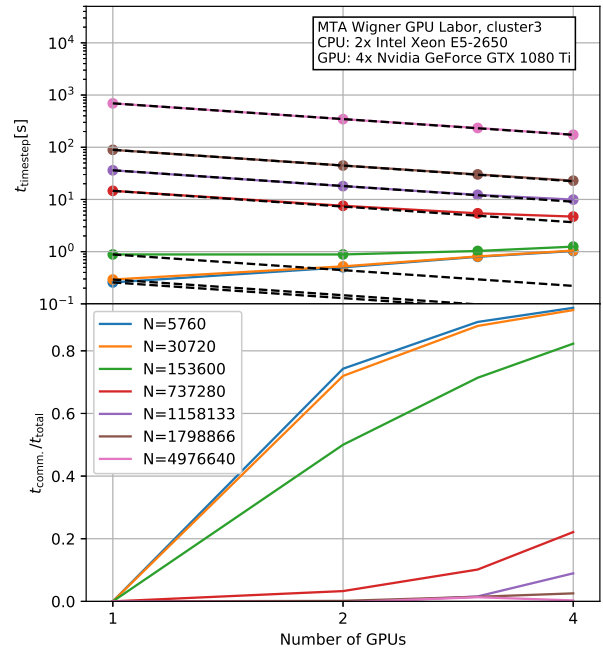}
        \caption{The efficiency of OpenMP-CUDA parallelization for the \StePS code. \textbf{Top:} The wall-clock time needed for one simulation timestep as a function of number of GPUs. The dashed line represents theoretical maximum of the achievable OpenMP parallelisation. \textbf{Bottom:} The ratio of the wall-clock time of the OpenMP communication, and of a full timestep. The different colors represent different particle numbers. See text for detailed discussion.}\label{fig:OpenMP_CUDA_scaling}
\end{figure}

\begin{figure}[h!]
    \centering
         \includegraphics[width=0.45\textwidth]{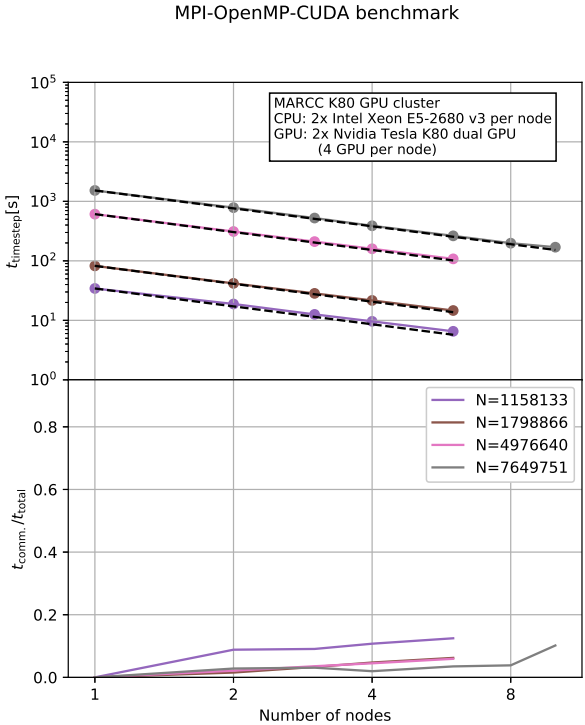}
        \caption{The efficiency of MPI-OpenMP-CUDA parallelization for the \StePS code. \textbf{Top:} The wall-clock time needed for calculating one timestep as a function of number of GPUs. The dashed line represents optimal MPI parallelization. \textbf{Bottom:} The ratio of the wall-clock time spent with communication between the threads, and of a full timestep. The different colors represent different particle numbers. See text for detailed discussion.}\label{fig:MPI_OpenMP_CUDA_scaling}
\end{figure}

We tested the effectiveness of the different parallelization methods with multiple non-periodic comoving $\Lambda$CDM cosmological simulations with different particle numbers in three different hardware settings. The first 10 time steps were timed directly in each case and for further scaling calculations we normalized the wall-clock time with the number of time steps for each simulation. We benchmarked the CPU-only MPI-OpenMP parallelization by running simulations with various particle and node number setups on the E\"otv\"os University (ELTE) Atlasz HPC2009 cluster\footnote{https://hpc.iig.elte.hu}. Note that this machine cannot be considered as a modern HPC, so rather the relative than the absolute values should be considered in the analysis. On Atlasz HPC2009 each node had two Intel Xeon E5520 CPUs, and OpenMP parallelization was used inside each computing node. For $N>10^6$, the communication cost was below 2\%, when 16 computing nodes were used. For testing the OpenMP-GPU paralellization, we run our simulations on a single computing node with four Nvidia GeForce GTX 1080 Ti GPUs in the GPU Laboratory of the Hungarian Academy of Sciences, the results can be seen in Fig.~\ref{fig:OpenMP_CUDA_scaling}. Our test simulation with $\sim7.37\cdot10^5$ particles achieved $\sim800$ times acceleration with just two GPUs with single precision, compared to one Xeon E5520 CPU node, used in the MPI-OpenMP benchmark. We have sustained 9.3Tflop/s on this node, and this is $\sim24\%$ of the theoretical peak performance and roughly $50\%$ of the efficiency of the $\varphi$-GPU direct N-body code\cite{Berczik_phiGPU_nbodypp6GPU}. Also, it is clear from this test, that using multiple GPUs is only worth it if the particle number is large enough. Our final test was done on the GPU cluster of the Maryland Advanced Research Computing Center\footnote{\url{https://www.marcc.jhu.edu/}} (MARCC). The measured timestep wall-clock times as a function of compute nodes can be seen in Fig.~\ref{fig:MPI_OpenMP_CUDA_scaling}. Up to 10 compute nodes with 20 Nvidia Tesla K80 dual-GPU cards were used, totalling 40 GPUs. The effectiveness of parallelization turned out to be over 93\% for our largest test run with $7.6\cdot10^6$ particles for 32 GPUs with 8 MPI tasks, above which the effectiveness started to decline.

\begin{table}[h]
	\centering
	\begin{tabular}{ | l | p{2cm} | p{2cm} | p{2cm} | }
	\hline
	$N_{GPU}$ & $t_{timestep} (s)$ & Number of MPI tasks & Efficiency \\ \hline
	1 & 5942.11 & 1 & 1.0 \\ \hline 
	2 & 3060.33 & 1 & 0.97083 \\ \hline
	4 & 1524.02 & 1 & 0.97474 \\ \hline
	6 & 1042.30 & 2 & 0.95016 \\ \hline
	8 & 784.11 & 2 & 0.94727 \\ \hline
	12 & 524.20 & 3 & 0.94463 \\ \hline
	16 & 388.65 & 4 & 0.95557 \\ \hline
	24 & 263.21 & 6 & 0.94065 \\ \hline
	32 & 198.10 & 8 & 0.93736 \\ \hline
	40 & 169.67 & 10 & 0.87554 \\
	\hline
	\end{tabular}
	\caption{Data from the MPI-OpenMP-CUDA hybrid parallelization test. The particle number was $7.6\cdot10^6$ in this test simulation. The $t_{timestep}(N_{GPU}=1)/(t_{timestep}(N_{GPU})\cdot N_{GPU})$ parallelization efficiency is above 87\% even for 40 GPUs. See text for discussion.}
\label{tab:EffPar}
\end{table}

\section{Generating Initial Conditions}
\label{sec:ICs}

The main motivation for running cosmological simulations is to calculate the evolution of statistical properties of the density field over time. To do this, one has to start the simulation from an initial particle distribution at early time with the right correlation function. The statistics of the density field at the epoch of the recombination ($z\simeq1100$) is known from the cosmic microwave background measurements\cite{2013ApJS..208...20B, 2018arXiv180706209P}. From this point, the power spectrum $P(k)$ as a function of wavenumber $k$ can be calculated for later times with perturbation theory. Since these analytic methods do not, or not fully take the nonlinear effects into account, perturbative techniques alone are not suitable for calculating the present $P(k)$ for small scales, yet they can be used to generate initial conditions (ICs) down to $z \approx 200$ with linear methods and $z \approx 50$ with second order techniques -- depending also on mass resolution\cite{2006MNRAS.373..369C}. In \cite{2018MNRAS.477.1949R}, we presented an IC generator algorithm that based on a remapping of an existing periodic initial condition with HEALPix\cite{2005ApJ...622..759G} tiling. Because the HEALPix tiling is not perfectly isotropic, the small artificial density fluctuations can grow and cause distortions during the simulation. In the rest of this section we present a new IC generation algorithm for \StePS simulations that is free from non-uniformity.

\subsection{Generating spherical glasses}

The first step of generating the initial conditions is to generate a particle distribution that represents a constant density field, and the net gravitational force acting on each particle is as small as possible. This is a non-trivial problem and it is not clear whether a correct answer even exists\cite{1989ApJS...70..419H,2005ApJ...634..728S}. Nevertheless, in case of periodic simulation with translational symmetry, two different solutions are used. The first method places the particles onto a three-dimensional grid, whereas the other solution uses a periodic glass. Using a periodic glass was first suggested by White\cite{1994astro.ph.10043W}, and it is used in most cosmological N-body simulations nowadays. Glasses are generated by placing point masses randomly in a periodic box, usually smaller than the simulation box itself, and evolving them in an Einstein-de Sitter universe with reversed gravity. Simulations with glassy initial conditions show significantly smaller discreteness artifacts at small scales compared to grid-based ICs. The other advantage of this approach is that glasses are more isotropic.

For a \StePS simulation, particle glasses have to be generated with radially decreasing resolution. We start by compactifying the real space using inverse stereographic projection which maps the 3 dimensional Euclidean space onto the 3D hypersurface of a 4D sphere. When Cartesian coordinates are used in the three-dimensional space, the transformation rules for the inverse stereographic projection are
\begin{equation}
        \begin{aligned}
        \omega &= 2 \cdot \arctan\left( \frac{\sqrt{x^2+y^2+z^2}}{D_s} \right)\\
        \vartheta &= \cos^{-1}\left( \frac{z}{\sqrt{x^2+y^2+z^2}} \right)\\
        \varphi &= \arctan\left( \frac{y}{x} \right)
        \end{aligned}
\label{eq:3Dster_proj_inv}
\end{equation}
where $\omega$, $\vartheta$ and $\varphi$ are the angular coordinates on the three-dimensional hypersurface of the four dimensional hypersphere, $D_s$ is the diameter of the hypersphere, and $x$, $y$, $z$ are the coordinates in real space. The forward transformations are given by
\begin{equation}
        \begin{aligned}
        x &= D_s \cdot \tan\left( \frac{\omega}{2} \right)\sin(\vartheta)\cos(\varphi)\\
        y &= D_s \cdot \tan\left( \frac{\omega}{2} \right)\sin(\vartheta)\sin(\varphi)\\
        z &= D_s \cdot \tan\left( \frac{\omega}{2} \right)\cos(\vartheta).
        \end{aligned}
\label{eq:3Dster_proj}
\end{equation}
We note that stereographic projection is only used when generating the initial conditions but all other calculations are done in real space to minimize floating point operations.

To generate initial conditions for \StePS simulations with radially decreasing resolution one starts by dividing the compactified space into slices along constant $\omega$ spheres. This is the equivalent of slicing the real space into concentric spherical shells. In every shell we place $N_{shell}$ particles randomly, and transform their coordinates into the real space with eq.~\ref{eq:3Dster_proj} stereographic projection. The masses of the particles in the shell with index $j$ are
\begin{equation}
	m_j = \rho_{\text{crit}}\cdot\Omega_m \cdot \frac{V_{\text{shell}, j}}{N_{\text{shell}}},
	\label{eq:StePSpartMass}
\end{equation}
where $V_{shell, j}$ is the real-space volume of the shell. The increase rate of particle mass with radius depends on the increase rate of volume of real space shells. Depending on the slicing of the compactified space, the increase rate can be controlled. We implemented two slicing schemes. 

\begin{figure}
	\includegraphics[width=\columnwidth]{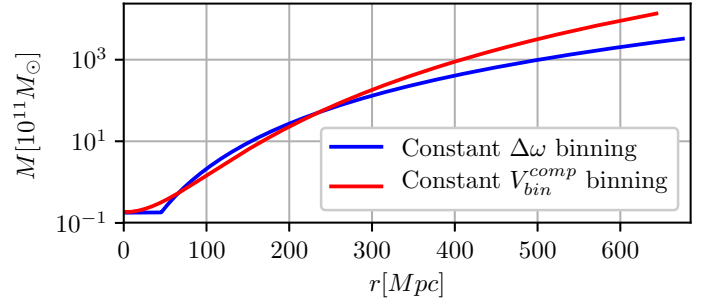}
	\caption{Initial mass resolution as a function of radius for the constant $\Delta\omega$ (blue curve) and the constant $V_{text{bin}}^{\text{comp}}$ (red curve) binning schemes for spherical glass generation. The parameters of the constant $\Delta\omega$ glass are the following: $D_s = 60.0$~Mpc, $r_c = 45.0$~Mpc, $D_{\text{sim}} = 684.9$~Mpc, $N_{\text{radial}}= 600$, $N_{\text{shell}} = 12288$. The constant $V_{text{bin}}^{\text{comp}}$ glass has the following parameters:  $D_s = 100.0$~Mpc, $D_{\text{sim}} = 684.9$~Mpc, $N_{\text{radial}}= 405$, $N_{\text{shell}} = 12288$.}
\label{fig:Resolution}
\end{figure}

The first scheme is called ``constant $\Delta \omega$ binning'', where the compactified space is divided into equally spaced shells along the $\omega$ compact coordinate but, to set a lower limit on particle mass, the innermost shells within an $\omega_c$ cut-off are united into a single volume with
\begin{equation}
	N_{\text{inner}} = \text{floor}\left(\frac{4\pi}{3}r_c^3\Omega_m\rho_{\text{crit}} / m_{\text{inner}} \right)
	\label{eq:StePSNinner}
\end{equation}
particles of equal mass, where $r_c$ is the real space radius corresponding to $\omega_c$ compactified coordinate. Since particle masses vary with radius, the introduction of the $\omega_c$ cut-off is important and the distinction between the innermost shells and outer shells is necessary to prevent the mixing of particles with too high mass ratios, which would cause artificial distortions in the simulation. The blue curve in Fig.~\ref{fig:Resolution} shows the particle mass as a function of radius where the unified innermost shells are visible as the constant line between $0 < r < r_c$. The main advantage of the ``constant $\Delta \omega$ binning'' method is that if one chooses $\omega_c$ large enough, the effect of different mass particle mixing will be minimal at the center.

The other implemented binning method is called the ``constant compact space volume'' method. In this case we define shells along the $\omega$ compact coordinate such way that the resulting shells have the same compactified volume. The equation for the lower limit of the $i$-th shell in the $\omega$ coordinate is
\begin{equation}
	\frac{8 V_{\text{bin}}^{\text{comp}} }{\pi D_s^3}\cdot i = 2\omega_i^l - \sin(2\omega_i^l),
	\label{eq:ConstCompOmegaBin}
\end{equation}
where $V_{\text{bin}}^{\text{comp}}$ is the volume of a bin in the compact space. This equation must be solved numerically for each shell. After the limits for the shells are calculated, we place $N_{shell}$ particles randomly into each shell, and transform the coordinates back to the real space. The particle masses are calculated from Eq.~\ref{eq:StePSpartMass} for every shell. With this method the particle masses will decrease smoothly outwards in the entire simulation volume, as it is shown by the red curve in Fig.~\ref{fig:Resolution}.

Many other realizations of space binnings are possible. Also, one can use different compactification maps or it is possible to change the angular resolution by setting $N_{\text{shell}}$ to $\omega$ dependent, etc. We will discuss these possibilities in a future study.

Once the spatial binning is defined and particle coordinates and masses are generated, we follow the standard way of glass generation. By integrating Eq.~\ref{eq:Comoving_Spherical_newtonian} with reversed gravity, after a sufficiently long relaxation period, the particles will settle down into a glass-like configuration. With the current implementation, glass generation takes a time comparable to running the simulation since in a spherical setting, one cannot use the trick of periodic glass generation with a significantly smaller box than the entire simulation volume. On the other hand, once a glass is generated, it can be reused for generating any number of initial conditions as long as particle number is the same.

\begin{figure*}
    \centering
	\begin{subfigure}[b]{0.42\textwidth}
		\includegraphics[width=\textwidth]{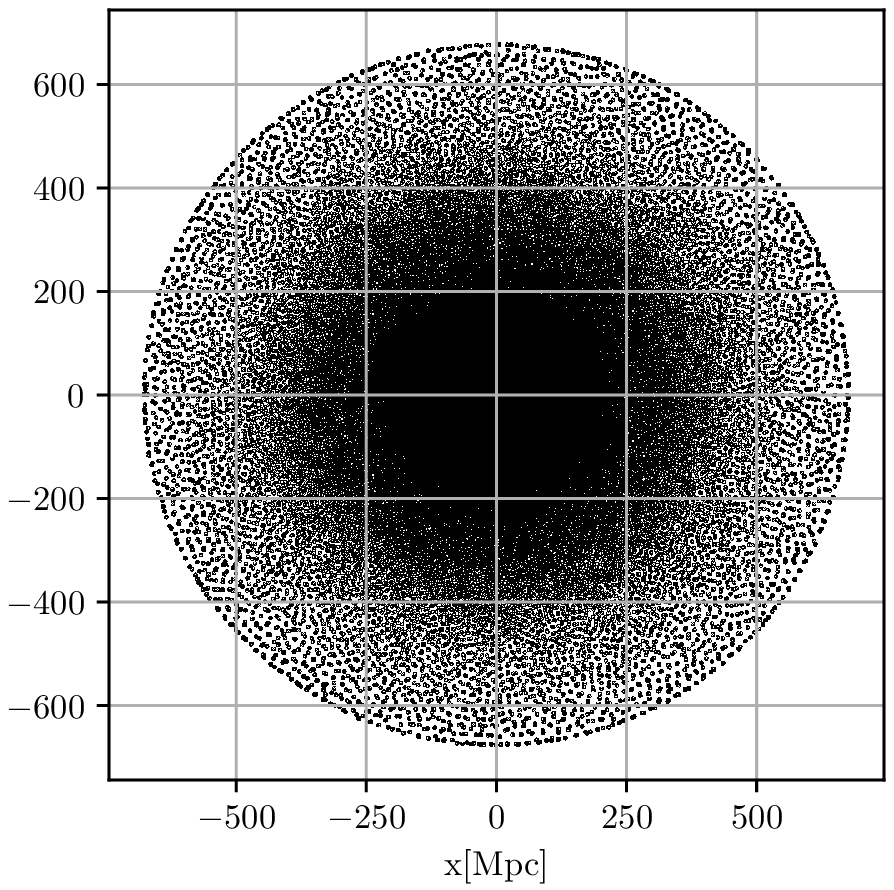}
	\end{subfigure}
	~
	\begin{subfigure}[b]{0.42\textwidth}
                \includegraphics[width=\textwidth]{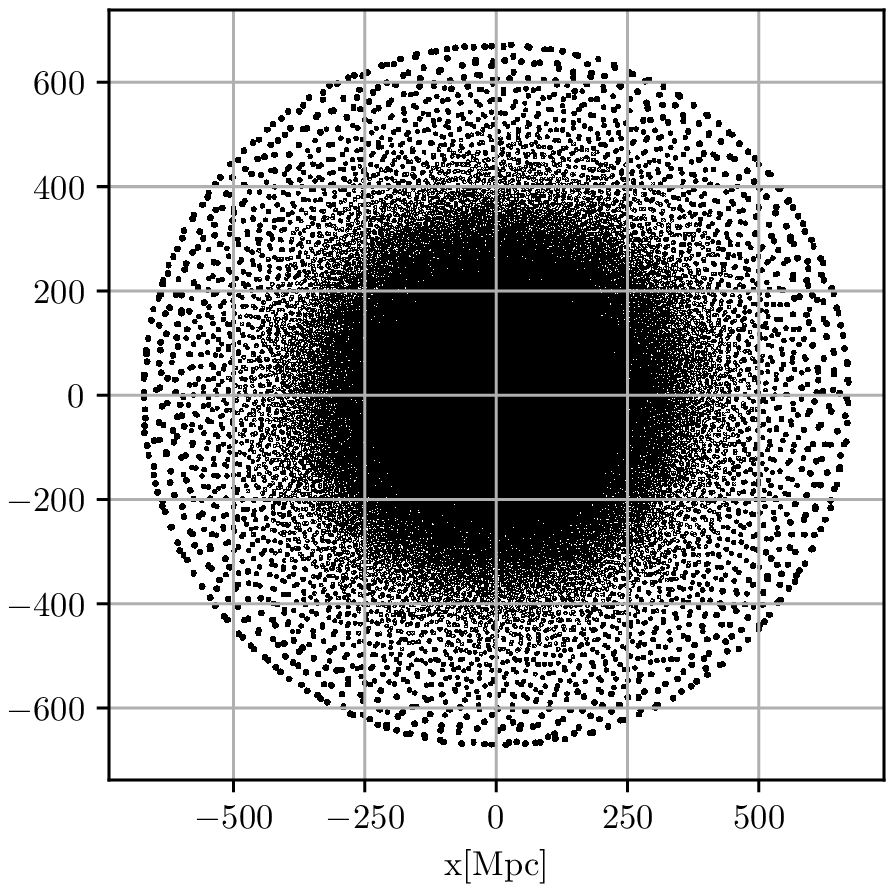}
        \end{subfigure}
		\caption{$4^\circ$ thick wedges cut out from Spherical glasses generated by the ``constant $\Delta \omega$ binning'' method (left panel) and the ``constant compact space volume'' method (right). The parameters of glass generation are the same as in the caption of Fig.~\ref{fig:Resolution}. We used an EdS Universe as a background throughout glass generation. The plotted size of the particles is proportional to their mass. The total number of the particles is $N=4.9\cdot10^{6}$ in both cases.}
\label{fig:GlassandRes}
\end{figure*}

We illustrate the generated glasses in Fig.~\ref{fig:GlassandRes} for both, ``constant $\Delta \omega$ binning'' and ``constant compact space volume'' methods for wedges of $4^\circ$ cut out of the glasses.

\subsection{Perturbing the glass particles}

Here we summarize the basic ideas of initial condition generation for periodic simulations, and introduce our new algorithm for the \StePS geometry.

The first step is to calculate the power spectrum for the initial time. In linear Eulerian perturbation theory, every $k$ mode of the $P(k)$ power spectrum is evolving independently, and modes can be scaled to any scale factor via the
\begin{equation}
	D(a) = \frac{5 \Omega_m H_0 }{ 2 } H(a) \int\limits^a_0 \frac{da'}{\dot{a'}^3}
\end{equation}
growth function\cite{2003moco.book.....D}. The power spectrum should be normalized at present day scale factor to be consistent with the $\sigma_8$ cosmological parameter. After the initial power spectrum is calculated, the $\delta(\mathbf{k})$ Fourier transform of the density fluctuation field can be generated assuming a Gaussian random field\cite{2005ApJ...634..728S}. 
$\delta(\mathbf{k})$ is calculated in a finite range: from zero to the $k_{Ny} = \pi/\Delta x$ Nyquist wavenumber in each dimension, which is given by the average particle spacing.

Once the density field is calculated, one only needs to perturb the positions and velocities of the particles to generate the initial conditions. In Lagrangian fluid dynamics, the movement of fluid elements is described by the $x(t_0)$ initial coordinates and the displacement field $\Psi(x(t_0), t)$. The perturbed coordinates become
\begin{equation}
	x(t) = x(t_0) + \Psi(x(t_0),t),
\end{equation}
where $t_0$ is the initial time, and $\Psi(x(t_0),t_0) = 0$. Lagrangian Perturbation Theory (LPT) uses a perturbative approach to calculate the displacement field from Fourier space density fluctuations \cite{1978IAUS...79..409Z, 2014MNRAS.439.3630W} in the form of
\begin{equation}
	\Psi(x(t_0),t) = \Psi^{(1)}(x(t_0),t) + \Psi^{(2)}(x(t_0),t) +  \Psi^{(3)}(x(t_0),t) + \cdots.
\end{equation}
The first order solution is called the Zel'dovich-approximation\cite{1970A&A.....5...84Z}, which can be written as 
\begin{align}
	\Psi^{(1)}(\mathbf{q}) &= \int \frac{i\mathbf{k}}{k^2} \delta(\mathbf{k}) e^{i\mathbf{q}\mathbf{k}}\\
		\mathbf{x} &= q + \Psi^{(1)}(\mathbf{q})\\
		\dot{\mathbf{x}} & = \frac{\dot{D}(a)}{D(1)}\Psi(\mathbf{q}), 
	\label{eq:Zeldovich}
\end{align}
where $\mathbf{q}$ are the initial, and $\mathbf{x}$ are the final coordinates of the particles. Of course, the $\Psi(q)$ displacement field is calculated in a cubic grid, and is interpolated to the original position of the glass particles. Similarly, the second order solution of Lagrange perturbation theory (2LPT)\cite{2006MNRAS.373..369C} is also used for initial condition generation. In case of periodic initial conditions with constant particle mass, particle masses are calculated directly from Eq.~\ref{eq:Const_part_masses}.

The IC generation algorithm for our spherically symmetric geometry is very similar but there are two main differences. The first difference is that we do not have a typical average particle spacing in the simulation volume because the mass and the spatial resolution decreases in the radial direction. The second difference is that we do not have a periodic geometry, so we can almost freely choose the box-size $L_{box}$ in which the fluctuations are calculated by using the Zel'dovich or 2LPT approximation. If one is interested in the effects of simulating sub-survey fluctuations only, $L_{box}/2 < R_{sim}$ should be chosen. In this case, the same density field will be repeated multiple times to cover the simulation sphere. On the other hand, when $L_{box}/2 > R_{sim}$ is chosen, super-survey modes can be simulated, although they will never be resolved and will be prone to cosmic variance due to the large but finite simulated volume. In this case the ICs will be truly non-periodic.

In traditional simulations, the cut-off at large $k$ is determined by the average particle separation which is proportional to the cubic root of particle number. In our case, average particle separation is not constant across the simulation volume but grows with the radial distance from the center. As the simplest solution, one can use the Nyquist wavenumber $k_{\textrm{Ny, inner}} = \pi/\Delta x_{\textrm{inner}}$ that corresponds to the best resolution at the center of the simulation. When the displacement field is applied to the particle glass, the lower resolution outer parts will undersample the density field which might lead to aliasing effects. In practice, however, these aliasing effects do not seem to affect simulation results. Despite of this, we implemented another method which is from undersampling. This second method works by computing the displacement field with a few different $k_{Ny}$ cutoff wavenumbers corresponding to the radius dependent $\Delta x(r)$ average displacements in spherical shells around the center of the simulation. The actual displacement of the particles is calculated by interpolating between the displacement fields with the nearest $k_{Ny}$ cutoffs. To calculate the displacement fields, we relied on the publicly available \texttt{NgenIC}\cite{2015ascl.soft02003S} and \texttt{2LPTic}\cite{2012ascl.soft01005C} codes for the Zel'dovich and the 2LPT approximation, respectively.

To be consistent to the desired cosmological parameters, the next step is to set the particle masses to
\begin{equation}
        m_j = m_j^{\text{glass}}\frac{\Omega_m\rho_{\text{crit}}\cdot\frac{4\pi}{3}R_{\text{sim}}^3}{\sum\limits^{N}_{i=1}m_i^{\text{glass}}},
\end{equation}
where $m_j$ is the mass of the $j$th particle, $m_j^{\text{glass}}$ is the original mass of the glass particle, and $R_{\text{sim}}$ is the radius of the simulation. If the goal is to generate ICs for a non-comoving simulation, one last step should be taken: rescaling the coordinates and adding the Hubble flow to velocities.

\section{Demonstration of the method}

For illustration, we aimed to re-simulate the Millennium Run\cite{2005Natur.435..629S} with the \StePS code. The Millennium Simulation had a great impact and demonstrated the effectiveness of the GADGET cosmological tree code. The original simulation had $2160^3$ particles in a periodic cube of $684.9\textnormal{Mpc}$ linear size. It used 512 processors and required about 28 days of wall-clock time on an IBM p690 supercomputer. 

We used the same cosmological parameters and random seed to generate the initial conditions for the non-periodic \StePS simulation. The radius of the simulation was set to $R_{sim} = 684.9\textnormal{Mpc}$, and we used constant $\Delta \omega$ binning, c.f. Sec.~\ref{sec:ICs}. The radius of the inner, constant resolution sphere around the simulation center was $r_c = 45$~Mpc. The spatial resolution at the center was $\sim 1.6$ times worse than the original Millennium Simulation. Our simulation ran for 106 wall-clock hours on 12 Nvidia GeForce 1080ti GPUs. The parameters of both simulations are summarized in Table~\ref{tab:SimParams}.

\begin{table}[h]
\begin{tabular}{ l | c c }
\hline
	& Millennium & StePS \\
\hline
	$\Omega_m$ & \multicolumn{2}{c}{$0.25$} \\
	$\Omega_\Lambda$ & \multicolumn{2}{c}{$0.75$} \\
	$H_0 \left[\textnormal{km/s/Mpc}\right]$ & \multicolumn{2}{c}{$73.0$}\\
	$\sigma_8$ & \multicolumn{2}{c}{$0.9$}\\
	Initial redshift & \multicolumn{2}{c}{$127$}\\

\hline
	linear size $\left[ \textnormal{Mpc} \right]$ & $L_\textnormal{box} = 684.9$ & $R_\textnormal{sim} = 684.9$ \\
	simulated volume $\left[ \textnormal{Gpc}^3 \right]$ & $0.321$ & $1.35$ \\
	number of particles & $1.01 \times 10^{10} $ & $1.17 \times 10^7$ \\
	particle mass $\left[ M_\odot \right]$ & $1.2 \times 10^{9}$ & $5.2\times10^{9}$-$10^{14}$ \\
force calculation & $\mathcal{O}(N \log N)$ & $\mathcal{O}(N^2)$ \\
memory use $\left[ \textnormal{GB} \right]$ & $\sim 1{,}000$ & $0.342$ \\
number of processing units & $512$ (CPU) & $12$ (GPU) \\
wall-clock time $\left[ h \right]$ & $683$ & $106$ \\
\hline
\end{tabular}
\caption{Main parameters of the Millennium Run and \StePS simulations.}
\label{tab:SimParams}
\end{table}

\subsection{Results}

\begin{figure*}
    \centering
        \begin{subfigure}[b]{1.0\textwidth}
                \includegraphics[width=\textwidth]{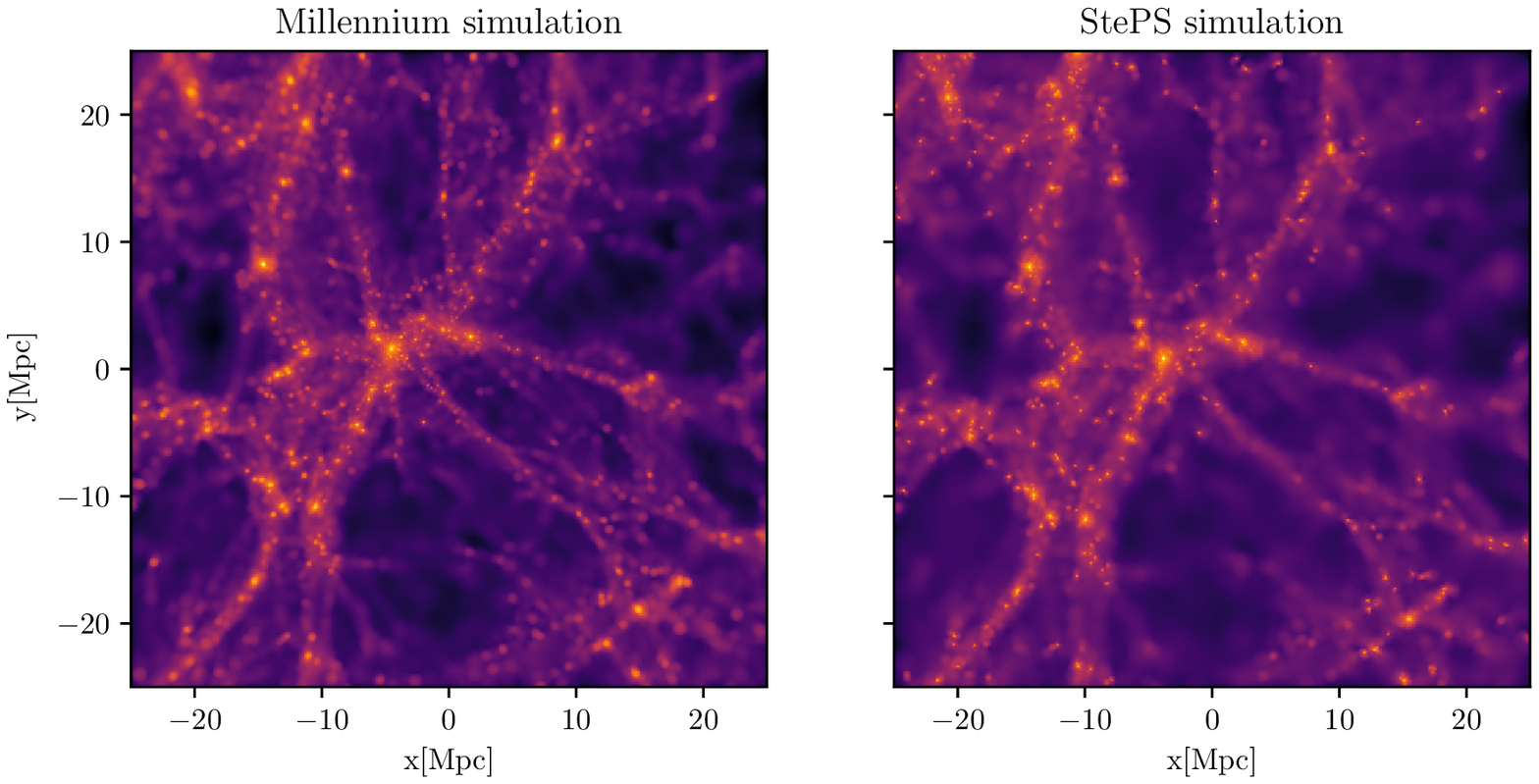}
        \end{subfigure}
        ~
        \begin{subfigure}[b]{1.0\textwidth}
                \includegraphics[width=\textwidth]{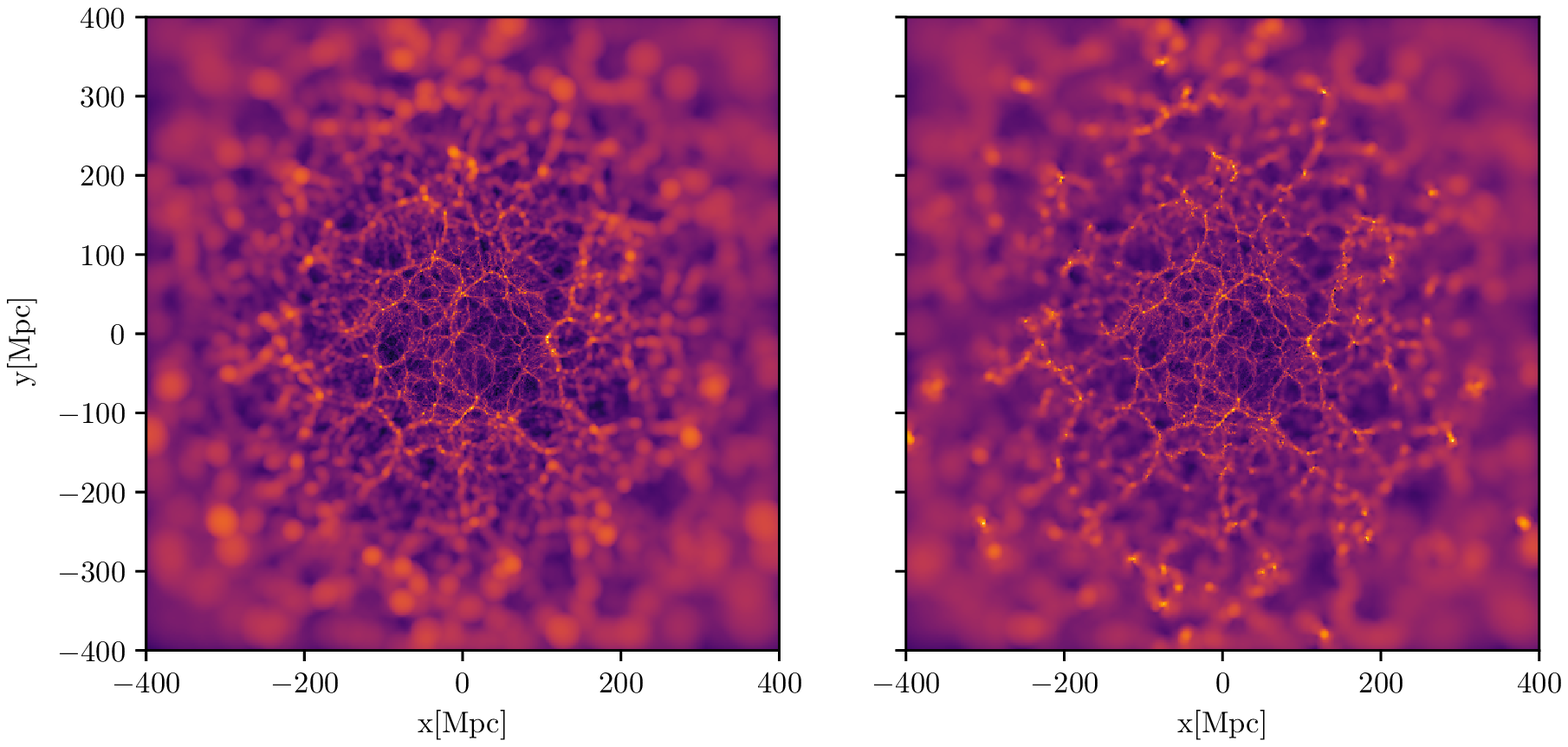}
        \end{subfigure}
		\caption{Visualization of distribution of the dark matter in the Millennium and in our \StePS simulation at $z=0$, in different scales. For the easier comparison we used a compactified version of the Millennium snapshot. The thickness of the density slices were $10Mpc$. See text for details.}
\label{fig:DensitySlices}
\end{figure*}

To compare the constant resolution Millennium Simulation to \StePS, we compactified the $z=0$ snapshot of the original Millennium Run by placing multiple copies of the simulation cube side by side and applying Eq.~\ref{eq:3Dster_proj_inv}. We aggregated the particles with constant $\Delta \omega$ binning in the radial direction and with equal-area HEALPix\cite{2005ApJ...622..759G} tiling in the $\vartheta$ and $\varphi$ coordinates. We averaged the positions, summed up the masses and inertia of the dark matter particles in each bin and substituted them with a single particle. After decompactification with Eq.~\ref{eq:3Dster_proj}, we arrived at a particle distribution with a very similar resolution to the \StePS simulation as a function of radius. The comparison of Millennium and the \StePS simulation can be seen in Fig.~\ref{fig:DensitySlices} at different scales. While the Millennium Simulation has slightly better resolution the the \StePS simulation at the very center (see the top panels of Fig.~\ref{fig:DensitySlices}), the structure at $z=0$ are identical. Some differences between Millennium and \StePS are visible at large distances from the center (see the bottom panels of Fig.~\ref{fig:DensitySlices}): structures in the \StePS simulation are sharper at large radii. This is due to the fact that compactification and time evolution are not interchangeable operations.

\section{Summary}

We presented a multi-GPU implementation of the \StePS code, and demonstrated the effectiveness of parallelization. We also implemented a new initial condition generator for compactified simulations. We were able to reproduce the Millennium Simulation at a spatial resolution slightly worse than the original at the very center of the \StePS simulations on 12 GPUs in under 106 hours. The source code of the simulation program and the IC generator script is freely available at our github repository (\url{https://github.com/eltevo/StePS}), licensed under GNU General Public License v2.0.

\section*{Acknowledgements}

The authors would like to thank Volker Springel and Adrian Jenkins for helping recreate the Millennium initial conditions. We also want to thank Robert Beck for stimulating discussions. This work has been supported by the NKFI grants NN 114560 and NN 129148. IS acknowledges support from National Science Foundation (NSF) award 1616974. We would like to thank the GPU Laboratory of the Hungarian Academy of Sciences, Wigner Research Centre of Physics for providing computing resources. Part of this research project was conducted using the computational resources at the Maryland Advanced Research Computing Center (MARCC).


\bibliography{StePS_MultiGPU_article}

\end{document}